\documentclass[11pt]{article}
\newcommand{\be}{\begin{equation}}
\newcommand{\ee}{\end{equation}}
\newcommand{\beq}{\begin{eqnarray}}
\newcommand{\eeq}{\end{eqnarray}}

\usepackage{graphicx}

\begin{document}
\begin{center}
{\bf \LARGE Non trivial generalizations of the Schwinger pair production 
result}\\
[7mm]
J. AVAN\footnote{Supported in part by a CNRS/Brown Accord }\\
{\em LPTHE, Tour 16 \\
Universit\'e Pierre et Marie Curie \\
75252 Paris Cedex 05 France}\\
[5mm]
H. M. FRIED\\
{\em Department of Physics \\
Brown University \\
Providence R.I. 02912 USA}\\
[5mm]
Y. GABELLINI\\
{\em Institut Non Lin\'eaire de Nice\\
 1361 Route des Lucioles\\
06560 Valbonne France}\\
[5mm]

\vspace{5mm}
Abstract
\end{center}

We present new, non trivial generalizations of the recent Tomaras, Tsamis and 
Woodard
extension of the original Schwinger formula for charged pair production in a 
constant electric field. That extension generalized the Schwinger result to 
electric fields 
$E_3(x_{\pm})$ dependent upon one $\underline{\rm or}$ the other light--cone
coordinates, $x_{+}$ or $x_{-}$, $x_{\pm} = x_{3} \pm x_{0}$; the present work 
generalizes their result to electric fields $E_3(x_{+}, x_{-})$ dependent upon 
$\underline{\rm both}$ coordinates. Displayed in the form of a final, functional 
integral, or equivalent linkage operation, our result does not appear to be 
exactly calculable in the general case; and we give a simple, approximate 
example when $E_3(x_{+}, x_{-})$ is a slowly varying function of its variables. 
We extend this result to the more general case where $\overrightarrow E$ can 
point in a varying direction, and where an arbitrary magnetic field 
$\overrightarrow B$ is present; both extensions can be cast into the form of 
Gaussian--weighted functional integrals over well  defined factors, which are 
amenable to approximations depending on the nature and variations of the fields.
	
\newpage

A recent, non trivial generalization of Schwinger's 1951 calculation of the 
probability/vol.sec for $e^+e^-$ production in a constant electric field has 
been given by Tomaras, Tsamis and Woodard ( TTW ) \cite{one} in which the 
electric field may depend upon either light--cone coordinates $x_{\pm} = x_{3} 
\pm x_{0}$, but not upon both. The form of their result is exactly the same as 
Schwinger's \cite{two}, which raises the question as to whether further 
generalizations are possible; and if so, what form they would take.

Shortly after the TTW paper appeared, a second and independent calculation of 
pair production in an electric field depending on either $x_{+}$ or $x_{-}$ was 
performed \cite{three} using functional techniques, and verifying the result of 
ref.\cite{one}. Those functional methods are here employed to attempt a further 
generalization to the case where the electric field depends upon both 
light--cone variables, $E = E_3(x_{+}, x_{-})$. The result is more complicated 
than the original Schwinger form, although the important ( and non trivial~)  
essential singularity in the region of small coupling is apparently preserved. 
The exact statement here requires the evaluation of an additional ( and non 
trivial~) functional integral; however, for certain situations, such as particle 
production in the overlap volume of a pair of high intensity lasers \cite{three} 
reasonable kinematic approximations are surely justified. If 
$$\displaystyle \ln P_0 = {\alpha E^2\over \pi^2}\sum_{n=1}^{\infty}{1\over 
n^2}\,e\,^{\displaystyle -n\pi m^2/gE}\ \ ,\ \ \ \ \alpha = {g^2\over 4\pi}$$ 
denotes the logarithmic density of the exact vacuum--persistence probability 
density of the Schwinger constant field calculation, the main result of this 
paper can be expressed as 
$$\displaystyle \ln P'_0 = {\alpha E^2\over \pi^2}\sum_{n=1}^{\infty}{1\over 
n^2}\,e\,^{\displaystyle -n\pi m^2/gE}{\cal M}_n$$
with $E = E(x_{+}, x_{-})$, $\tau_n = n\pi/gE$,  $J(\lambda_1, \lambda_2) = 
\theta(\lambda_1-\lambda_2)\lambda_2+\theta(\lambda_2-\lambda_1)\lambda_1-
\lambda_1\lambda_2$, and 
\beq
{}&{\cal M}_n= (-1)^n\, e\,^{\displaystyle 
4m^2\tau_n\int\!\!\!\int_0^1\!d\lambda_1d\lambda_2\,{\delta\over \delta 
v_{+}(\lambda_1)}\,J(\lambda_1, \lambda_2)\,{\delta\over \delta 
v_{-}(\lambda_2)}}\nonumber\\
&\times\displaystyle 
\cos\biggl[g\,\tau_n\int_0^1\!\!d\lambda\,E\Bigl(x_+-m^{-1}v_{+}(\lambda), 
x_--m^{-1}v_{-}(\lambda)\Bigr)\biggr]\Biggl\vert_{v_{\pm}\to 0}
\eeq
where the `` linkage operation '' of $(1)$ may be cast into a corresponding 
functional integral. It doesn't seem to be possible to evaluate $(1)$ exactly, 
although kinematic approximations are certainly possible. One of these will be 
used below to illustrate the formula. Note that if $E$ depends on only one 
variable $x_+$  or $x_-$, ${\cal M}_n\to 1$, and the TTW extension of the 
Schwinger result is obtained.

We hold to the notation and technique of ref.\cite{three}, beginning with the 
exact statement of the vacuum persistence amplitude in the absence of radiative 
corrections
\be<\!0\,|\,S\,|\,0\!> = e\,^{\displaystyle L[A_{ext}]}\ee
The vacuum persistence probability of the present problem, here called $P'_0$, 
is then given by $P'_0 = \exp\Bigl[2\,{\rm Re}\,L[A_{ext}]\Bigr]$, and the 
probability of producing (~at least ) one charged pair is $P'_1 = 1 - P'_0$. 
Note that $L[A]$ is really $L[F]$, since $L$ is rigorously gauge invariant; but 
we shall continue to use the vector--potential description, beginning with the 
exact Fradkin representation \cite{four} 
\beq
{}&L[A] = -\displaystyle{1\over 2}\int_0^{\infty}\!{ds\over 
s}\,e\,^{\displaystyle{-ism^2}}\!\!\exp\,\biggl[i\int_0^s\!ds'\sum_{\mu}{\delta^
2\over\delta v_{\mu}^2(s')}\biggr]\nonumber\\
&\times\displaystyle\int\!{d^4p\over(2\pi)^4}\,\,e\,^{\displaystyle{ip\!\cdot\!\
!\!\int_0^s\!ds'v(s')}}\displaystyle  \int \!d^4\!x\,e\,^{\displaystyle 
-ig\int_0^s\!ds'v_{\mu}(s') \,A_{\mu}(x-\int_0^{s'}\!\!v)}\\
&\times\,{\rm tr}\biggl(\displaystyle e\,^{\displaystyle 
g\!\int_0^s\!ds'\sigma\!\cdot\!F(x-\int_0^{s'}\!\!v)}\biggr)_+\Bigg\vert_{\,v_{\
mu}\rightarrow 0}\!-(g\rightarrow 0)\nonumber
\eeq
The functional linkage operation of $(3)$ may be recast into that of a 
Gaussian--weighted functional integral, since 
$$e\,^{\displaystyle i\int_0^s\!ds\,{\delta^2\over\delta v^2(s')}}\,{\cal 
F}[v]\bigg\vert_{\,v\rightarrow 0} = N\!\!\int\!\! d[v]\, e\,^{\displaystyle 
{i\over 4}\int_0^s\!ds\,v^2(s')}\,{\cal F}[v]$$
for arbitrary ${\cal F}[v]$, with the normalization $N$ given by $$N^{-1} = 
\displaystyle\int\!\! d[v]\, \exp\biggl[{\displaystyle {i\over 
4}\int_0^s\!ds\,v^2(s')}\biggr]$$
The linkage formulation, which dispenses with normalization constants, is 
somewhat more convenient, and will be followed whenever possible.

To represent a given, external field in the $\hat 3$ direction, we may choose 
$A_{\mu} = ( \vec A_{\perp}, A_3, A_0 ) \to ( 0, A_3, A_0 ) \to ( 0, A_+, A_- )$ 
with $\vec A_{\perp} = 0$, $A_{\pm}(x) = \sum_{\mu}\,n_{\pm}^{\mu} A_{\mu}(x_+, 
x_-)$. Here, $n_{\pm}^{\mu} = (0, 0, 1, \mp 1)$, so that  $n_{\pm}\!\cdot \!a = 
a_{\pm} = a_3 \pm a_0$. Note that $n_+^2 = n_-^2 = 0$, $n_+\!\cdot\! n_- = 2$, 
and that $p\!\cdot\!A\to p_3A_3 - p_0A_0 = {1\over 2}( p_+A_- + p_-A_+)$. We 
shall choose the gauge specified by $A_- = 0$, so that $A^2 = A_{\perp}^2 + 
A_+A_- = 0$, and $E(x_+, x_-) = -\displaystyle{\partial A_+\over\partial 
x_+}(x_+, x_-)$.

As in ref.\cite{three}, we extract the $v_{\mu}(s'')$ dependence inside the 
arguments of $A_{\mu}$ and $F_{\mu\nu}$, or of $A_+$ and $E$, by introducing a 
functional form of unity into $(3)$, replacing ${\cal 
F}\displaystyle\Bigl[\int_0^{s'}\!\!v_+(s''),\int_0^{s'}\!\!v_-(s'')\Bigr]$ by 
\be
\displaystyle\int\!\! d[u_+]\!\int\!\! d[u_-]\, {\cal F}\Bigl[ u_+(s'), 
u_-(s')\Bigr]\,\delta\Bigl[ u_+(s') - 
\int_0^{s'}\!\!v_+(s'')\Bigr]\,\delta\Bigl[ u_-(s') - 
\int_0^{s'}\!\!v_-(s'')\Bigr]
\ee
where the $\delta$--functional notation means that when the region $0$ -- $s$ is 
broken up into many small intervals labeled by the discrete indices $s_i$, $i = 
1,\ldots , N$ ( with $N\to\infty$, subsequently ), 
$\displaystyle\delta\Bigl[Q(s')\Bigr]\to \prod_{i=1}^N\delta\Bigl(Q(s_i)\Bigr)$.

Note that since we are treating each $v_{\pm}(s')$ as a continuous ( although 
not necessarily differentiable ) function of $s'$, the $u_{\pm}(s')$ introduced 
in $(4)$ have, at least, a continuous first derivative.

Each $\delta$ function of $(4)$, and the overall $\delta$ functional, may be 
written in terms of a standard Fourier representation; so that, in the 
continuous limit, $(4)$ may be replaced by 
\beq
{}&\displaystyle N'^2\,\int\!\! d[u_+]\!\int\!\! d[u_-]\, {\cal F}\Bigl[ 
u_+(s'), u_-(s')\Bigr]\,\int\!\! d[\Omega_+]\!\int\!\! d[\Omega_-]\\
&\hskip-1truecm\times\,e\,^{\displaystyle 
i\int_0^s\!ds'\Bigl[u_+(s')\Omega_+(s') +  
u_-(s')\Omega_-(s'\Bigr]}\,e\,^{\displaystyle 
-i\int_0^s\!ds'\Bigl[\Omega_+(s')\int_0^{s'}\!\!v_+(s'') +  
\Omega_-(s')\int_0^{s'}\!\!v_-(s'')\Bigr]}\nonumber
\eeq
where $N'$ is an appropriate normalization factor ( which disappears from the 
final result ). With the aid of Abel's trick, the last exponential factor of 
$(5)$ may be written as 
$$
\displaystyle\exp\biggl\{ 
-i\int_0^s\!ds'\,v_{\mu}(s')\Bigl[n_+^{\mu}\!\int_{s'}^{s}\!\!ds''\,\Omega_+(s''
) +  n_-^{\mu}\!\int_{s'}^{s}\!\!ds''\,\Omega_-(s'')\Bigr]\biggr\}
$$
so that the entire $v$--linkage operation is immediate : 
\beq
{}&\hskip-0.5truecm e\,^{\displaystyle i\int_0^s\!ds\,{\delta^2\over\delta 
v^2(s')}}e\,^{\displaystyle i\int_0^s\!ds'v_{\mu}(s')\Bigl[p_{\mu} - g 
A_{\mu}(s') - n_+^{\mu}\!\int_{s'}^{s}\!\!ds''\,\Omega_+(s'') -  
n_-^{\mu}\!\int_{s'}^{s}\!\!ds''\,\Omega_-(s'')\Bigr]}\bigg\vert_{\,v\rightarrow 
0}\nonumber\\
&\displaystyle = \exp\Bigl\{-i\int_0^s\!ds'\Bigl[p_{\mu} - g A_{\mu}(s') - 
n_+^{\mu}\!\int_{s'}^{s}\,\Omega_+ -  
n_-^{\mu}\!\int_{s'}^{s}\Omega_-\Bigr]^2\Bigl\}
\eeq
where we have used the notation $A_{\mu}(s')\equiv A_{\mu}(x_+ - u_+(s'), x_- - 
u_-(s'))$.
As in ref.\cite{three}, the explicitly quadratic $\Omega_+$ and $\Omega_-$ terms 
of the exponential factor of $(6)$ are removed because $n_+^2 = n_-^2 = 0$, but 
there remains a non zero $\Omega_+\!\cdot\!\Omega_-$ cross term, so that $(6)$ 
becomes
\beq
{}&\displaystyle\exp\Bigl\{-isp^2 +ig\,p_{-}\!\int_0^s\!ds' \,A_{+}(s') + 
2i\,p_{+}\!\int_0^s\!ds' \,s' \,\Omega_{+}(s') + 2i\,p_{-}\!\int_0^s\!ds' \,s' 
\,\Omega_{-}(s')\nonumber\\
&\displaystyle -2i\,g\!\int_0^s\!ds' \,\Omega_{+}(s')\!\int_0^{s'}\!ds'' 
\,A_{+}(s'') - 
4i\!\int_0^s\!ds_1\!\int_0^s\!ds_2\,\Omega_{+}(s_1)h(s_1,s_2)\Omega_{-}(s_2)
\Bigr\}
\eeq
where $h(s_1,s_2) = \theta (s_1-s_2)s_2 +  \theta (s_2-s_1)s_1 = {1\over 2} ( 
s_1 + s_2 - \vert s_1-s_2\vert)$ is that function of proper time which always 
appears when constructing representations of $G_c[A]$  \cite{five}. In writing 
$(7)$, we have used the `` inverse '' form of Abel's manipulation.

Performing the $\int\!{d^4p}$, one requires
\beq
&\displaystyle\int\!{d^4p}\, e\,^{\displaystyle-isp^2 + ip\!\cdot\!\Bigl[n_- g 
\!\int_0^s\!A(s') + 2 n_+ \int_0^s\!s'\,\Omega_+(s') + 2 n_- 
\int_0^s\!s'\,\Omega_-(s')\Bigr]}\nonumber\\
&\hskip-0.5truecm\displaystyle = -i\Bigl({\pi^2\over 
s^2}\Bigr)\exp\Bigl\{4i\!\int\!\!\!\int_0^s\! 
\!ds_1ds_2\,\Omega_+(s_1)\Bigl({s_1s_2\over s}\Bigr)\Omega_-(s_2) + {2i\over 
s}\int_0^s\!s'\,\Omega_+(s')g\! \!\int_0^s\!\!A_+(s')\Bigr\}
\eeq
where we have again used the properties : $n_+^2 = n_-^2 = 0$, $n_+\!\cdot\! n_- 
= 2$. The new $\Omega_+\Omega_-$ term of $(8)$ may be combined with the 
previous, like term of $(7)$, forming the combination 
$\displaystyle\exp\Bigl\{-4i\!\!\int\!\!\!\int_0^s\!\Omega_+J\Omega_-\Bigr\}$, 
$J(s_1,s_2) = h(s_1,s_2)- \displaystyle{s_1s_2\over s}$; and the functional 
integral over $\Omega_{\pm}$ may be written as 
\beq
&\displaystyle e\,^{\displaystyle 4i\!\int\!\!{\delta\over \delta 
u_{+}(s_1)}\,J(s_1, s_2)\,{\delta\over \delta u_{-}(s_2)}}\,N'^2\!\int\!\! 
d[\Omega_+]\!\int\!\! d[\Omega_-]\,e\,^{\displaystyle 
i\!\int_0^s\!\!ds'u_-(s')\Omega_-(s')}\nonumber\\
& \displaystyle\times\, e\,^{\displaystyle 
i\!\int_0^s\!\!ds'\Omega_+(s')\Bigl[u_+(s') + 2g{s'\over s}\int_0^{s}\!\!ds'' 
A_+(s'') - 2g\int_0^{s'}\!\!ds'' A_+(s'')\Bigr]}\\
&\displaystyle = e\,^{\displaystyle 4i\!\int\!\!{\delta\over \delta 
u_{+}}\,J\,{\delta\over \delta u_{-}}}\delta\Bigl[ u_-(s')\Bigr]\,\delta\Bigl[ 
u_+(s') + 2g{s'\over s} \int_0^{s}\!\!A_+(s'') - 
2g\int_0^{s'}\!\!A_+(s'')\Bigr]\nonumber
\eeq
Eq.$(9)$ is multiplied by the $u_{\pm}$--dependent factor
\be
\displaystyle {\rm tr} \biggl(e\,^{\displaystyle 
g\!\int_0^s\!\!ds'\,\sigma\!\cdot\!F\Big(x_+-u_+(s'),x_--u_-(s')\Bigr)}\biggr)_+ 
= 4\cosh\Bigl(g\!\int_0^s\!\!ds'E(s')\Bigr)
\ee
where $E(s')\equiv E\Big(x_+-u_+(s'),x_--u_-(s')\Bigr)$. The 
$\displaystyle\int\!\! d[u_+]\!\int\!\! d[u_-]$ may be evaluated by transferring 
( by an integration by parts over each $u$ variable~) the differential 
$\displaystyle{\delta/\delta u_{\pm}}$ operators which act upon the $\delta$ 
functionals of $(9)$ to equivalent operations upon the $u_{\pm}$ dependence of 
$(10)$
$$
\displaystyle4\Big\vert\det{\delta u\over\delta f}\Bigr\vert \, 
e\,^{\displaystyle 4i\!\int\!\!{\delta\over \delta u_{+}}\,J\,{\delta\over 
\delta u_{-}}}\cosh\Bigl(g\!\int_0^s\!\!ds'E(s')\Bigr)\bigg\vert_{u_-(s') = 
0,u_+(s') = u(s')}
$$
where $\displaystyle\Big\vert\det{\delta u\over\delta f}\Bigr\vert$ is the 
determinant of the transformation from the variables $u_+(s')$ to the variables 
$f(s')$, the latter given by the argument of the $\delta$ functional produced by 
the $\displaystyle\int\!\! d[\Omega_+]$. Here, $u(s')$ is the solution of the 
integral equation
\be
u(s') = 2g\Bigl[\int_0^{s'}\!\!ds''\,A_+(s'') - {s'\over s} 
\int_0^{s}\!\!ds''\,A_+(s'')\Bigr]
\ee
with $u_-(s') = 0$, corresponding to the other $\delta$ functional of $(9)$.

By exactly the same arguments as in \cite{three}, the only possible solution to 
$(11)$ is $u(s') = u'(s') = 0$, which means that, as in  \cite{three},
\be
\displaystyle\Big\vert\det{\delta u\over\delta f}\Bigr\vert = {gsE(x_+,x_-)\over 
\sinh(gsE(x_+,x_-))}
\ee
 with the difference that this $E$ can depend upon both $x_+$ and $x_-$.

The sign of $E$ entering into $(10)$ and $(12)$ is irrelevant, and we shall 
suppose it is positive. Our expression for $L[A]$ can therefore be put into a 
form which closely resembles that of Schwinger, and of ref.\cite{three}.
\beq
&\hskip-1truecm\displaystyle L[A] = 
{i\over8\pi^2}\!\!\int\!\!d^4x\!\int_0^{\infty}\!{ds\over 
s^2}\,e\,^{\displaystyle{-ism^2}}{gE\over \sinh(gsE)}\,e\,^{\displaystyle 
4ism^2\!\!\int\!\!\!\!\int_0^1\!\!d\lambda_1d\lambda_2{\delta\over \delta 
v_{+}(\lambda_1)}\,J(\lambda_1, \lambda_2)\,{\delta\over \delta 
v_{-}(\lambda_2)}}\nonumber\\
&\hskip-2truecm 
\displaystyle\times\cosh\Bigl[gs\int_0^{1}\!d\lambda\,E(x_+-m^{-1}v_+(\lambda),x
_--m^{-1}v_-(\lambda))\Bigr]\bigg\vert_{v_{\pm}\to0} - ( g\to 0 )
\eeq
In writing $(13)$, we have observed that, in the interval $0$ -- $s$, any 
continuous function $u_{\pm}(s')$ may be given by a Fourier series 
representation as a function of $(s'/s)$, and we therefore replace $u_{\pm}(s')$ 
by $m^{-1}v_{\pm}(s'/s)\equiv m^{-1}v_{\pm}(\lambda)$, where $\lambda = s'/s$, 
and the factor $m^{-1}$ is inserted for dimensional reasons (~and will cancel 
away when the functional derivatives $\delta/ \delta v_{\pm}$ are taken ). 
However, because 
$$
{\delta u_{\pm}(s_1)\over\delta u_{\pm}(s_2)} = \delta(s_1-s_2) = {1\over s 
}\delta(\lambda_1-\lambda_2)\hskip0.3truecm,\ {\rm and}\ \ {\delta 
v_{\pm}(\lambda_1)\over\delta v_{\pm}(\lambda_2)} = \delta(\lambda_1-\lambda_2)
$$
one must adopt the relation $\displaystyle{\delta\over\delta u_{\pm}(s')} = 
{m\over s}{\delta\over\delta v_{\pm}(\lambda')}$. After extracting an overall 
factor of $s$, $J(\lambda_1,\lambda_2)$ is understood to be given by 
$$J(\lambda_1,\lambda_2) = \theta(\lambda_1-\lambda_2)\lambda_2 +  
\theta(\lambda_2-\lambda_1)\lambda_1 - \lambda_1\lambda_2$$

Extracting the ${\rm Re}\,L[A]$ from an expression involving an operator such as 
that of $(13)$ requires some further thought, and we therefore imagine that the 
linkage operator of $(13)$  is expanded in powers of $J$, so that the rotation 
of contour $s\to -i\tau$ is permitted; here, that expansion yields powers of 
$\displaystyle4m^2\tau\!\!\int\!\!\!\!\int_0^1\!\!{\delta\over \delta 
v_{+}(\lambda_1)}\,J(\lambda_1, \lambda_2)\,{\delta\over \delta 
v_{-}(\lambda_2)}$, which generate  real results because $E$ and all of its 
derivative are real. Under this contour rotation, one again finds that the 
contributions to ${\rm Re}\,L[A]$ arise from zeroes of the denominator 
$\sin(g\tau E)$, occuring for $\tau\to\tau_n - i\varepsilon$, $\tau_n = 
n\pi/gE$; and we then sum up all $J$--dependent terms, so that
$$2{\rm Re}\,L[A] = -{\alpha\over\pi^2}\!\!\int\!\!d^4x\,E^2 
\sum_{n=1}^{\infty}{1\over n^2}\,e\,^{\displaystyle -n\pi m^2/gE}{\cal 
M}_n(x_+,x_-)$$
with ${\cal M}_n$ defined in $(1)$. This seems to be as far as the exact 
analysis can be performed.

Without approximation, the linkage operation of $(13)$ may be converted to a 
pair of functional integrals, 
\beq
&\hskip-1truecm\displaystyle e\,^{\displaystyle iab\!\int\!\!{\delta\over \delta 
v_{+}}\,J\,{\delta\over \delta v_{-}}}\,{\cal 
F}[v_{+},v_{-}]\Bigl\vert_{v_{\pm}\to0}\nonumber\\
&\hskip-2truecm \displaystyle = 
\det[J]\int\!d[\chi_+]\!\int\!d[\chi_-]\,e\,^{\displaystyle 
i\!\int\!\!\chi_+\,J\,\chi_-}\,{\cal F}\Bigl[a\!\int\!\!J\,\chi_+, 
b\!\int\!\!J\,\chi_-\Bigr]
\eeq
and mean field/stationary phase approximations derived for the right hand side 
of $(14)$. Perhaps the simplest approximation of all occurs when $E(x_+,x_-) = 
E_+(x_+) + E_-(x_-)$, a form suggested by recent estimates of pair production in 
the overlap region of two, crossed, high intensity lasers \cite{three}. If all 
the derivatives of $E$ of order higher than the second are neglected, \beq
&\displaystyle\int_0^1\!\!d\lambda 
\,E_{\pm}(x_{\pm}-m^{-1}v_{\pm}(\lambda))\simeq E_{\pm}(x_{\pm}) - 
m^{-1}\!\!\int_0^1\!\!d\lambda \,v_{\pm}(\lambda){\partial E_{\pm}(x_{\pm})\over 
 \partial x_{\pm}}\nonumber\\
&\displaystyle +{m^{-2}\over 2}\!\!\int_0^1\!\!d\lambda 
\,v_{\pm}^2(\lambda){\partial^2 E_{\pm}(x_{\pm})\over  \partial x_{\pm}^2} + 
\cdots\nonumber
\eeq
which approximation assumes that the fractional variations of the electric field 
are very small over distances on the order of the Compton wavelenght.

Then, it is easy to see that the linkage operation yields
\be
\hskip-0.5truecm{\cal M}_n = \cos\Bigl[{a^2\over 2}Q(\alpha_+^2\gamma_- + 
\alpha_-^2\gamma_+)\Bigr]\exp\Bigl[-4a^2\alpha_+\alpha_-R-{1\over 2}{\rm 
Tr}\ln(1+a^2\gamma_+\gamma_-I)\Bigr]
\ee
where : $a\!=\!4\pi n/gE$, $\displaystyle\alpha_{\pm}\!=\!n\pi {\partial \over  
\partial x_{\pm}}\ln\Bigl(E/m^2\Bigr)$, 
$\displaystyle\gamma_{\pm}\!=\!{n\pi\over E} {\partial^2 \over  \partial 
x_{\pm}^2}E$, $E\!=\!E_+ + E_-$
and : 
$$
Q = \int\!\!\!\!\int_0^1\!\!d\lambda_1d\lambda_2<\!\lambda_1\,\vert \,I\left({ 
1+a^2\gamma_+\gamma_-I}\right)^{-1}\vert\,\lambda_2\!>
$$
$$
R = \int\!\!\!\!\int_0^1\!\!d\lambda_1d\lambda_2<\!\lambda_1\,\vert \,J\left({ 
1+a^2\gamma_+\gamma_-I}\right)^{-1}\vert\,\lambda_2\!>
$$
with
$$
I(\lambda_1,\lambda_2) = \,<\!\lambda_1\,\vert \,I\vert\,\lambda_2\!>\, = 
\int_0^1\!\!d\lambda\, J(\lambda_1,\lambda) J(\lambda,\lambda_2) 
$$
Note that, formally, for large $n$, $R\sim n^{-4}$ while $a^2x_+x_-\sim n^{4}$, 
so that the exponential terms of $(15)$ cannot grow rapidly with $n$. One 
therefore infers that the essential singularity structure of the original 
Schwinger result is preserved.

For an electric field $E_3$, which depends upon $x_{\perp}$, one can retain the 
condition $A_- = A_{\perp} = 0$, but require $A_+$ to depend on $x_{\perp}$, as 
well as $x_+$ and $x_-$. One then introduces variables $u_{\perp}(s')$ and 
$\Omega_{\perp}(s')$, and following the above analysis finds
\beq
&\hskip-1truecm\displaystyle L[A] = {i\over32\pi^2}\!\!\int_0^{\infty}\!{ds\over 
s^3}\,e\,^{\displaystyle{-ism^2}}\!\!\int \!\!d^4x\!\!\int \!\!d[u_+]\!\int 
\!\!d[u_-]\!\int\!\! d[u_{\perp}]\,{\rm tr}\biggl(e\,^{\displaystyle 
g\!\!\int_0^s\!\!\sigma\!\cdot\!F}\biggr)_+\nonumber\\
&\hskip-1.5truecm\displaystyle\times\,e\,^{\displaystyle 
4i\!\!\int\!\!{\delta\over \delta u_{+}}\,J\,{\delta\over \delta u_{-}} + 
i\!\!\int\!\!{\delta\over \delta u_{\perp}}\,J\,{\delta\over \delta 
u_{\perp}}}\,\delta\Bigl[ u_+(s') + 2g{s'\over s} \int_0^{s}\!\!A_+(s'') - 
2g\int_0^{s'}\!\!A_+(s'')\Bigr]\\
&\hskip-1truecm\displaystyle\times\,\delta\Bigl[ u_-(s')\Bigr]\delta\Bigl[ 
u_{\perp}(s')\Bigr]\nonumber
\eeq
An integration by part on the $u_{\pm}$, $u_{\perp}$ dependence then allows the 
linkage operator to act upon the ${\rm tr}\biggl(e\,^{\displaystyle 
g\!\!\int_0^s\!\!\sigma\!\cdot\!F}\biggr)_+$ term only; and one sees that the 
jacobian of the transformation between the $u_{\pm}(s')$ and $f(s')$ variables 
is exactly the same as in $(12)$, except that here $E = E_3(x_+, x_-, 
x_{\perp})$. If we restrict $\vec E$ to lie in the $3$ direction, then again,  
${\rm tr}\biggl(e\,^{\displaystyle 
g\!\!\int_0^s\!\!\sigma\!\cdot\!F}\biggr)_+\to 4\cosh(gsE)$ and the limits 
$u_{\pm} = u_{\perp} = 0$ are to be taken after the linkage operator of $(16)$ 
acts upon the $u_{\pm}$, $u_{\perp}$ dependence inside $4\cosh(gsE)$, with $E = 
E_3(x_+-u_+(s'), x_--u_-(s'), x_{\perp}-u_{\perp}(s'))$.

For arbitrary $\vec E$ and $\vec B$ fields, one must include at the very 
beginning $A_{\perp}(x_+, x_-, x_{\perp}) \not = 0$, although it is still 
convenient to retain the gauge condition $A_-=0$. Now there will appear in this 
final generalization of $(16)$ the extra factor
$$
{\cal F}[u] = \exp\Bigl[-ig^2 \int_0^{s}\!\!ds''A_{\perp}^2(s'') + i\,{g^2\over 
s}\Bigl(\int_0^{s}\!\!ds''A_{\perp}(s'')\Bigr)^2\Bigr]
$$
where,  following the notation of $(6)$, $A_{\perp}(s') = A_{\perp}(x_+-u_+(s'), 
x_--u_-(s'), x_{\perp}-u_{\perp}(s'))$, and ${\cal F}[u]$ multiplies the trace 
of the now more complicated ordered exponential ( OE ) 
$\biggl(e\,^{\displaystyle g\!\!\int_0^s\!\!\sigma\!\cdot\!F}\biggr)_+$. 
Following techniques developed for unitary OEs \cite{six}, the present OE can be 
approximately evaluated analytically when $\vec E$ and $\vec B$ directions 
change in ways that can be characterized as slowly varying ( `` adiabatic '' ) 
or rapidly fluctuating ( `` stochastic '' ); and the same linkage operator as in 
$(16)$  is then to act upon the product ${\cal F}[u]\biggl(e\,^{\displaystyle 
g\!\!\int_0^s\!\!\sigma\!\cdot\!F}\biggr)_+$. Note that there will appear 
another jacobian, involved in the variable change from $u_{\perp}(s')$ to 
$f_{\perp}(s')$, as in $(12)$, and both jacobians will simultaneously involve 
$\vec E$ and $\vec B$. Alternatively, if one wishes to avoid the direct 
evaluation of these jacobians, one can rewrite this generalization of $(16)$ in 
terms of functional integrals, and attempt to use mean field/stationary phase 
methods for their approximate evaluation. This extension of $(16)$ provides a 
functional representation of $L[A]$ appropriate for the most general choice of 
$\vec E$ and $\vec B$ fields. When the latter are constants, and in particular 
for $\vec B = 0$, it reduces immediately to Schwinger's 1951 solution, while it 
generates a definition of $L[A]$ at any stage inbetween.

Finally, a more symmetric formulation, explicitly depending only on the 
$F_{\mu\nu}$ may be obtained by starting from formula (5.22) of the book of 
ref.\cite{four}  
\beq
&\hskip-1truecm\displaystyle L[F] = -{1\over2}\!\!\int_0^{\infty}\!{ds\over 
s}\,e\,^{\displaystyle{-ism^2}}\!\!\int \!\!d^4x\!\!\int 
\!\!{d^4p\over(2\pi)^4}\,e\,^{\displaystyle i\int_0^{s}\!{\delta^2\over\delta 
v^2}}\,e\,^{\displaystyle ip\!\cdot\!\!\!\int_0^{s}\!v}\nonumber\\
&\hskip-1truecm\displaystyle\times\, N'\!\!\int \!\!d[\Omega]\!\int 
\!\!d[u]\,e\,^{\displaystyle 
i\!\!\int_0^{s}\!\!u_{\mu}\Omega_{\mu}}\,e\,^{\displaystyle 
-i\!\!\int_0^{s}\!\!\Omega_{\mu}\!\int_0^{s'}\!\!\!\!v_{\mu}}\\
&\hskip-1truecm\displaystyle\times\,e\,^{\displaystyle 
-ig\!\!\int_0^{s}\!\!ds'\,v_{\mu}(s')u_{\nu}(s')\!\int_0^{1}\!\!\lambda d\lambda 
\,F_{\mu\nu}(x-\lambda u(s'))}{\rm tr}\biggl(e\,^{\displaystyle 
g\!\!\int_0^s\!\!\sigma\!\cdot\!F(x-u(s'))}\biggr)_+-(g\to 0)\nonumber
\eeq
where every factor of $\displaystyle\int_0^{s'}\!\!ds''v_{\mu}(s'')$ has been 
replaced by $u_{\mu}(s')$. Performing the linkage operation, $\displaystyle\int 
\!\!{d^4p}$ and $\displaystyle\int \!\!d[\Omega]$, then leads to the result
\beq
&\hskip-1truecm\displaystyle L[F] = {i\over32\pi^2}\!\!\int_0^{\infty}\!{ds\over 
s^3}\,e\,^{\displaystyle{-ism^2}}\!\!\int \!\!d^4x\!\prod_{\mu=1}^4\!\int 
\!\!d[u]\,\Delta[u_{\mu}]\nonumber\\
&\hskip-1truecm\displaystyle\times\, e\,^{\displaystyle 
i\!\int\!{\delta\over\delta u_{\mu}}J{\delta\over\delta u_{\mu}}}\,\,{\rm 
tr}\biggl(e\,^{\displaystyle 
g\!\!\int_0^s\!\!\sigma\!\cdot\!F}\biggr)_+\,e\,^{\displaystyle Q}\ ,\\
&\hskip-1truecm\displaystyle \Delta[u_{\mu}]\ = \delta\Bigl[u_{\mu}(s') - 2g 
\int_0^{s}\!\!ds''(\theta(s'-s) - {s'\over s})\int_0^{1}\!\!\lambda d\lambda 
\,F_{\mu\nu}(x-\lambda u(s''))u_{\nu}(s'')\Bigr]\nonumber\\
&\hskip-1truecm\displaystyle Q = 
-ig^2\!\!\int_0^{s}\!\!ds'\!\int_0^{1}\!\!\lambda 
d\lambda\!\int_0^{1}\!\!\lambda' 
d\lambda'\!\int_0^{s}\!\!ds''u_{\nu}(s')u_{\sigma}(s'')\,F_{\mu\nu}(x-\lambda 
u(s'))F_{\mu\sigma}(x-\lambda' u(s''))\Bigl[\delta(s'-s'') - {1\over 
s}\Bigr]\nonumber
\eeq
The only solutions for each $u_{\mu}(s')$ allowed by the product of the four 
$\delta$ functional of $(18)$ are $u_{\mu}(s')\equiv 0$, although the jacobians 
obtained from each variable change $u_{\mu}(s')\to f_{\mu}(s')$ will each be 
unity only for the Schwinger case of constant  $F_{\mu\nu}$; in this latter 
situation, the $\exp\Bigl[{\displaystyle i\!\int\!{\delta\over\delta 
u_{\mu}}J{\delta\over\delta u_{\mu}}}\Bigl]$ operation may be recast into that 
of soluble gaussian functional integrals over the $u_{\mu}(s')$, leading back to 
Schwinger's original result.


\vskip2cm


\begin{thebibliography}{**}

\bibitem{one} T.N. Tomaras, N.C. Tsamis and R.P. Woodard, Phys. Rev. D {\bf62}, 
125005 (2000) 
\bibitem{two} J. Schwinger, Phys. Rev. {\bf82}, 664 (1951). See also L.S. Brown 
and T.W.B. Kibble, Phys. Rev. {\bf133}, 705 (1954); and the many references 
listed therein.
\bibitem{three} H.M. Fried, Y. Gabellini, B.H.J. McKellar and J. Avan, Phys. 
Rev. D {\bf63}, 125001 (2001)
\bibitem{four} E.S. Fradkin, Nucl. Phys. {\bf76}, 588 (1966). A detailed 
explanation and derivation of the Fradkin representation can be found in the 
book by H.M. Fried, Functional Methods and Eikonal Models, Editions 
Fronti\`eres, Gif sur Yvette, France (1990).
\bibitem{five} H.M. Fried and Y. Gabellini, Phys. Rev. D{\bf51}, 890 (1995) and 
H.M. Fried and Y. Gabellini, Phys. D{\bf51}, 906 (1995)
\bibitem{six} M.-E. Brachet and H.M. Fried, Phys. Lett. A{\bf103}, 309 (1984) 
and J. Math. Phys. {\bf28}, 15 (1987); H.M. Fried, J. Math. Phys. {\bf28}, 1275 
(1987) and J. Math. Phys. {\bf30}, 1161 (1989)
\end{thebibliography}
\end{document}